\documentclass[prb,aps,twocolumn,draft,tightenlines,%
showpacs,floatfix]{revtex4}  

\usepackage{graphicx}
\usepackage{dcolumn}
\usepackage{bm}
\usepackage{psfig}
\begin{document}

\title{Spin filtering through a double-bend structure}
\author{Q. W. Shi}
\author{J. Zhou}
\author{M. W. Wu}
\thanks{Author to whom correspondence should be addressed}%
\email{mwwu@ustc.edu.cn}%
\affiliation{Hefei National Laboratory for Physical Sciences at
  Microscale, University of Science and
Technology of China,  Hefei, Anhui, 230026, China and\\
Department of Physics, University of Science and
Technology of China, Hefei, Anhui, 230026, China}
\altaffiliation{Mailing Address.}

\date{\today}

\begin{abstract}
We propose a simple scheme for the spin filter by studying the
coherent transport of electrons through a double-bend structure
in a quantum wire with a weak lateral magnetic potential which
is much weaker than the Fermi energy of the leads. {\em Extremely
large} spin polarized current in the order of {\em micro-Ampere}  
can be obtained because of the strong resonant behavior 
from the double bends. 
Further study suggests the roubustness of this spin filter.
\end{abstract}
\maketitle

The rapid emerging field of spintronics promises to provide
 new advances that have a substantial impact 
on future applications.\cite{prinz,loss,spin}
Effective and efficient electrical spin injection of  spin-polarized 
current into semiconductors is  one of the major challenges 
in this field.\cite{wolf,mon,filip}
One method is to inject spin current through ideal 
ferromagnet/semiconductor 
interface. However, the polarization of the injected current 
is rather small due to 
the large conductivity mismatch.\cite{schmidt}  
The use of spin filters 
is therefore an alternative approach which can significantly enhance 
spin injection efficiencies. In some previous works, 
spin-selective barriers \cite{gilbert}
or stubs \cite{stub} are essential to realize spin polarization (SP). Other 
methods such as quantum dot\cite{fran} and resonant tunneling 
diode \cite{koga} also have been reported.  Very recently we also
proposed a scheme of spin filter by utilizing the ``band-gap''
generated by the weak lateral magnetic modulations.\cite{wu}
However, it is noted that the spin currents of these filters
are relatively small.
 
In this letter we propose a new scheme of the spin filter 
which provides {\em exteramly large} spin current
by utilizing the resonance in 
a double-bend structure with a uniform small magnetic field
which can be realized  by sticking a magnetic strip 
on top of the sample or using magnetic semiconductor.
The effect of the bend discontinuity has been discussed in detail 
in a mode-matching theory by Weisshaar {\sl et al.}\cite{wei}
There it was shown that strong resonance effects are 
present in the transmission 
coefficient versus energy due to the presence of a perpendicular 
single right bend. They further showed that the effect of 
the second bend ({\em i.e.} a double-bend structure)
is to add further fine resonances  superimposed on the dominant 
resonance, with the width and the spacing in energy depending
on the cavity of length $L$. We will show that this resonance
effect can be effectively utilized to generate SP's.

A schematic of the double-bend structure is shown in Fig.\ 1.
The spin dependent potential with Zeeman-like form
 $V_\sigma(x,y)=\sigma V_0g(x,y)$ is applied on the double bends
(regions B and C).
Here $g(x,y)=1$ if $(x,y)$ locates at regions B and C , and 
0 otherwise.  $\sigma$ is $\pm1$ for spin-up and -down electrons,
respectively. $V_0$ denotes a spin-independent parameter for 
the strength of the potential.
For $E_f\gg V_0$, spin-up and -down electrons experience different 
potentials: the spin-up electrons coherently transport through
a ``transparent'' barrier while the spin-down ones do
through a well. Therefore, spin polarized current can be obtained 
because of the mismatch of the resonances from the double bends 
of the spin-up and 
-down electrons.  

\begin{figure}[htb]
\vskip-0.1cm
\centerline{\psfig{figure=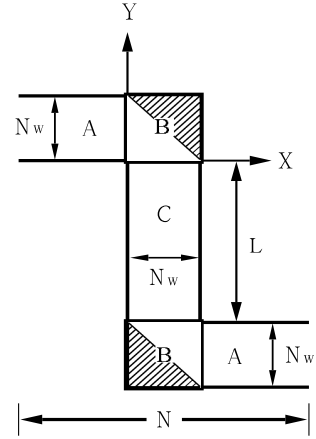,width=4cm,height=6cm,angle=0}}
\vskip-0.3cm
  \caption{Schematic of the spin filter in double-bend structure.}
\end{figure}

We describe the double-bend structure by a tight-binding 
Hamiltonian with the nearest-neighbour
approximation:
\begin{eqnarray}
H&=&\sum_{l,m,\sigma}(\epsilon_{l,m,\sigma} c^\dagger_{l,m,\sigma}
c_{l,m,\sigma} + 
t_{0}c^\dagger_{l,m+1,\sigma}c_{l,m,\sigma} \nonumber \\
&&\mbox{}+ t_{0}c^\dagger_{l+1,m,\sigma}c_{l,m,\sigma}) +\mbox{H.C.}\ ,
\end{eqnarray}
in which $l$ and $m$ denote the  coordinates along the  $x$- and $y$-axis
respectively.  
$\epsilon_{l,m,\sigma}=\epsilon_0+\sigma V_0$ ($=\epsilon_0$)
when $(l,m)$ locates at the B and C regions (when $(l,m)$ locates at
the A region), denotes the on-site energy
with $\epsilon_0=-4t_{0}$.
 $t_{0}=-\hbar^2/2m^\ast a^2$ is the hopping energy with $m^\ast$ 
and $a$ standing for the effective mass and  the ``lattice'' constant
respectively. 

The spin dependent conductance is calculated using  the
Landauer-B\"{u}ttiker\cite{Bu} formula with the help of 
the Green's function method.\cite{Da} The two-terminal spin-resolved 
conductance is given by
$G^{\sigma \sigma^\prime}=(e^2/h)\mbox{Tr}[\Gamma^{\sigma}_{1}
G^{\sigma\sigma^\prime+}_{1N}
\Gamma^{\sigma^\prime}_{N}G^{\sigma
^\prime\sigma -}_{N1}]$ with  $\Gamma_1$ 
($\Gamma_{N}$) representing the
self-energy  function for the isolated ideal leads.\cite{Da} 
We choose the  perfect ideal ohmic contact between the
leads and the semiconductor. $G^{\sigma\sigma^\prime+}_{1N}$ and 
$G^{\sigma\sigma^\prime-}_{N1}$ are the 
retarded and advanced Green functions
for the conductor, but with the effect of the leads included. 
The trace is performed over the spatial degrees of freedom along the 
$y$-axis. The spin dependent current within an energy window
$[E,E+\Delta E]$ is given by $I_\sigma=\int_{E}^{E+\Delta
  E}G^{\sigma\sigma}(E)dE$.

We perform a numerical calculation for a quantum wire with
 width $N_w$. A hard wall potential is applied in this
transverse direction which makes the lowest energy of the
$n$th subband (mode) be $\epsilon_n(N_w)=2|t_0|\{1-\cos[n\pi/(N_w+1)]\}$.
$a=9.53$\ {\AA} which makes $|t_0|=1$\ eV throughout the computation.
We take the Zeeman splitting energy $V_0=0.01|t_0|$. 

\begin{figure}[htb]
\vskip-0.3cm
 \centering{\psfig{figure=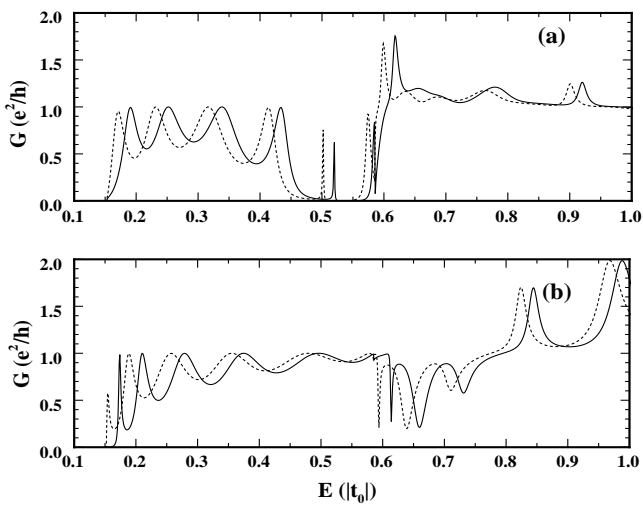,width=9cm,height=8cm,angle=0}} 
\vskip-0.3cm 
\caption{Conductance versus the energy of the electron 
for (a) the double-bend structure in Fig.\ 1 and (b)
the same structure in (a) but with the corners (shadowed areas
in Fig.\ 1)  cut. $L=20a$. Solid curve:
 $G^{\uparrow\uparrow}$; Dotted curve: $G^{\downarrow\downarrow}$. 
}
\end{figure}

In Fig.\ 2(a) the conductance is plotted as a function 
of the Fermi energy $E$ 
of the leads with $N_w=7a$ and cavity length $L=20a$. 
Both the single and the double modes are included in the
figure.
It is seen from the figure that  SP is obtained from each 
energy window in which the mismatch of the resonances 
for electrons with different spin directions occurs.  
Spin current densities  can be
obtained from the energy window $[0.51|t_0|,0.525|t_0|]$ with 
$I^{SP}_{\uparrow}=I_{\uparrow}-I_{\downarrow}\approx 61.7$\ nA  for 
spin-up current and from the window $[0.49|t_0|,0.51|t_0|]$ with
$I^{SP}_{\downarrow}=I_{\downarrow}-I_{\uparrow}\approx 63.7$\ nA 
for spin-down current, each with 100\ \% SP. SP can also be obtained
from other energy intervals due to the mismatch of the resonance
peaks for different spin. Particularly if one
chooses the energy window $[0.425|t_0|,0.49|t_0|]$, 
one gets an {\em extremely large} spin current
$I^{SP}_{\uparrow}\approx 0.635$\ $\mu$A. 
This large spin current  scales with the magnitude of the 
applied potential $V_0$. If one takes an even smaller number
$V_0=0.005|t_0|$ ($0.001|t_0|$), one also gets a large current 
$I^{SP}_{\uparrow}\approx 0.347$\ $\mu$A (0.072\ $\mu$A) 
by choosing a suitable energy window on the 
edge of the gap. 

The spin-independent gap near $E=0.55|t_0|$
corresponds to the anti-resonance gap 
due to the reflection of the bend structure. 
By cutting off the corners (the shadowed areas in Region B shown 
in Fig.\ 1) from the both bends, one can see from Fig.\ 2(b) 
that the gap disappears and one also loses the energy window for
the large spin current.

In order to understand the resonance feature of the double bends,
we calculate the conductance with $L=10a$ and $L=30a$. By comparing 
Fig.\ 3 and Fig.\ 2(a), one finds 
that the number of the resonant peaks increases with the cavity length $L$.
When the wave length 
of the incident electron $\lambda=2\pi a\sqrt{\frac{|t_0|}{E-\epsilon_1(N_w)}}$ 
satisfies the standing wave condition $\frac{j}{2}\lambda=L+2N_w$ 
with $(j=1,\cdot\cdot\cdot,j_{max})$, the conductance reaches the maximum. 
It is therefore easy to see that within the fixed
energy interval of the first subband ($0.15|t_0|<E<0.56|t_0|$),
a larger bend distance $L$ corresponds to a larger $j_{max}$ and therefore
more resonance peaks.

 \begin{figure}[htb]
\vskip-0.3cm
\centering{\psfig{figure=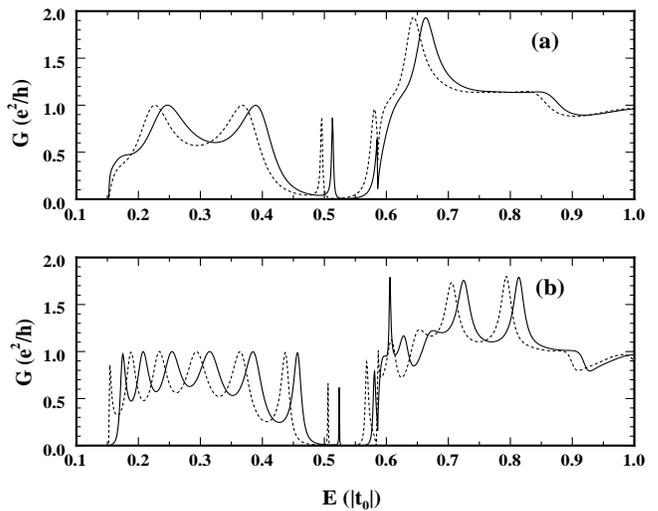,width=9cm,height=8cm,angle=0}}
\vskip-0.3cm
\caption{Conductance versus the energy of the electron with different $L$: (a) $L=10a$; (b) $L=30a$. 
 Solid curve: $G^{\uparrow\uparrow}$; Dotted curve: $G^{\downarrow\downarrow}$. }
\vskip-0.3cm
\end{figure}
   
We also check the roubustness of the spin filter proposed 
above by including Anderson disorder 
to investigate its effect on the spin polarized currents. 
We take the strength of the disorder to 
be $W=0.05|t_0|$, five times of the potential $V_0$.  We still obtain a 
large spin polarized current  $I^{SP}_{\uparrow}\approx 0.545$\ $\mu$A.

Finally we point out that the scheme of the spin filter proposed in
this letter can be generalized to any structure in which
the  conductance oscillates with the energy of the incident electrons.
Then by applying a small spin-dependent potential, one gets the mismatch
of the resonance peaks for different spins and hence the SP.
However, if one wishes to obtain a filter which can give a large spin 
current, then a anti-resonance gap is essential in the structure. 
 
In summary, we have proposed a simple scheme for spin filter 
by studying the coherent transport through double-bend structure with
a lateral magnetic potential. Extremely large spin current is
predicted from this structure.
The magnetic potential can be realized by sticking the magnetic 
strip on top of the sample or using magnetic semiconductors.
This spin filter is very robust to the disorder. 

One of the authors (MWW) was supported by the  ``100 Person Project'' 
of Chinese Academy of Sciences and Natural Science Foundation of China 
under Grant Nos. 9030312 and 10247002.

\end{document}